# On The Optimality Of All-To-All Broadcast In k-ary n-dimensional Tori

Jean-Pierre Jung[1], Ibrahima Sakho[2]

*UFR MIM, Université de Metz*
*Ile du Saulcy BP 80794 - 57012 Metz Cedex 01 – France*
[1]jpjung@univ-metz.fr
[2]sakho@univ-metz.fr

*Abstract*—All-to-all broadcast is a collective communication in a network with the constraint that every node must send to each other certain piece of its data. This paper addresses the problem of optimal all-port all-to-all broadcast in multidimensional tori. The optimality criteria considered are the minimum exchange steps, no duplicated data in the sense that only new data are conveyed to receivers and the balance of the communication links-load. It is proved that under these constraints, an optimal broadcast is not feasible in any multidimensional torus. Then, the tori which are capable of optimal broadcasts are characterized.

## I. INTRODUCTION

Parallel computers with distributed memory constitute an attractive alternative in the search of scalable architectures for massively parallel applications. Given the processors interconnection network, IN for short, of such computers, inter-processor communications, IPC for short, are realized by passing messages. Then intensive IPC can rapidly result in a bottleneck for the IN. In order to insure efficient IPC, several IN have been proposed in the literature. Among them, cartesian product graphs which generalize multidimensional meshes and tori are more popular.

Among communication patterns that induce intensive IPC, collective communication as defined in [1], [2] has received considerable attention. Collective communication is a communication pattern where a group of processors has to exchange data. Commonly used collective communication is the one where the group of processors is constituted of all the processors. Examples of such communication are all-to-all personalized communication [3], [4] and all-to-all broadcast [5], [6]. While in all-to-all personalized communication each node has to send a distinct message to every other node, in all-to-all broadcast each node has to send the same message to all others nodes. They are undoubtedly the most demanding for IN bandwidth and then the most demanding for execution time.

All-to-all broadcast is important in numerous applications that include protocols required for the control of distributed execution and intensive computation. Examples of such protocols are decentralised consensus [7], coordination of distributed checkpoints [8] and acquisition of new global state of a system [9]. Examples of intensive computation are sorting [10] and ordinary differential equation solving [11].

Performance models of all-to-all broadcast are generally based on parameters as the number of data exchange steps, the size of data exchanged at each step and the so-called NODUP in [19] which imposes the absence of redundancy that is every data convey only new information to its receiver.

It is obvious that any k-ary n-dimensional torus can not realise optimal all-to-all broadcast under all these constraints. The aim of this paper is then to characterize k-ary n-dimensional tori capable to realise such optimal all-to-all broadcasts.

The remainder of the paper is organized in five sections. Section II presents the related works and Section III the context of the study. Section IV presents mathematical properties used in the next sections to characterize k-ary n-cube tori suitable for optimal all-to-all broadcast and Section V the characterization of such tori. Section VI concludes the paper and presents the perspectives for future works.

## II. RELATED WORKS

Beyond the works cited in Section I, several studies have been conducted to devise efficient all-to-all broadcast algorithms for multidimensional meshes and tori. They can be classified in two main classes of algorithms: the direct algorithms like in [12] and [13] and the message combining algorithms like in [14], [15] and [16].

Direct algorithms aim at the minimisation of the number of data exchanges steps and then suppose that every pair of processors can directly exchange data. They then do not take into account the distance between the processors.

Message combining algorithms are more realistic. They aim at the minimisation of the size of the data exchanged at each step. Then data destined for a processor are combined in successive exchanges steps to result in longer data and a reduced start up cost.

Beyond these algorithms, there are other ones based on data pipelining techniques as described in [17] and [18].

In [20], the more general problem of methodologies for devising optimal all-to-all algorithms is addressed; an optimal all-to-all broadcast algorithm is proposed for k-ary 2-dimensional tori. The constraints of such a broadcast are:
- to route data on the shortest paths,

- to balance the link loads,
- to receive each data piece one and only once.

## III. PROBLEM DESCRIPTION

This section deals with the formulation of the optimal all-to-all broadcast problem. Definitions and properties of multidimensional tori which are essential for this formulation will be given.

### A. Definitions

*1) k-ary n-dimensional torus:* A k-ary n-dimensional torus is a network of $k^n$ nodes $x(x_1, x_2, \ldots, x_i, \ldots, x_n)$ such as $0 \leq x_i \leq k-1$ and two nodes are connected if and only if their addresses differ by 1 [modulo k] on one and only one digit.

More formally, it is a cartesian product of n rings having k nodes each one. Fig. 1 illustrates a bi-directional 5-ary 3-dimensional torus.

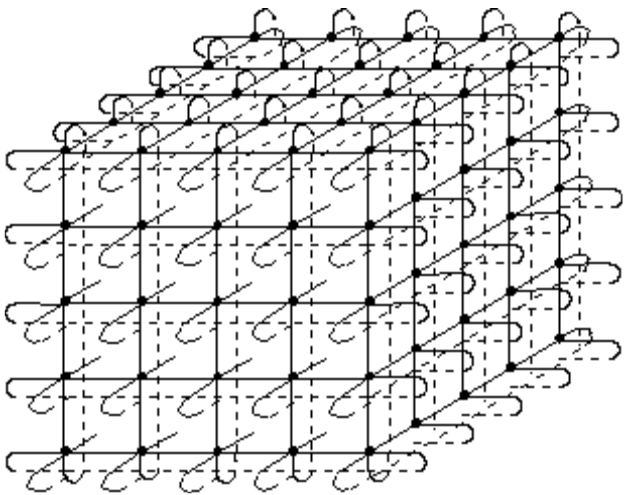

Fig. 1. Bidirectional 5-ary 3-dimensional torus

*2) All-to-all broadcast:* An all-to-all broadcast is a type of global communication in which each node has to broadcast the same atomic data to the others nodes.

At the end of such a communication, each node must be in possession of the data sent by all the nodes in the network.

Let T be the all-to-all broadcast time function.

*3) Optimal all-to-all broadcast:* An optimal all-to-all broadcast is the one realized within the shortest time. More formally, an all-to-all broadcast $A^*$ is optimal if and only if $T(A^*) \leq T(A)$ for any all-to-all broadcast A.

*4) ε-optimal all-to-all broadcast:* An ε-optimal all-to-all broadcast is the one realized within the shortest realisable time. More formally, an all-to-all broadcast $A^*$ is ε-optimal if and only if $T(A^*) < T(A) + \varepsilon$ for any all-to-all broadcast A.

### B. Properties of k-ary n-cube tori

The following properties come from the structure of the torus.

**Property 1:** Let d(x, y) be the distance between two nodes x and y. $d(x, y) = \sum_{1 \leq i \leq n} \text{Min}(|x_i - y_i|, k - |x_i - y_i|)$.

**Property 2:** The diameter of a k-ary n-dimensional torus is equal to $n \lfloor k/2 \rfloor$ where $\lfloor r \rfloor$ stands for the floor of r.

From Definitions 2-4, we can deduce the following characterization of optimal broadcasts.

**Proposition 1:** A necessary and sufficient condition for an all-to-all broadcast to be optimal is that:
a) each piece of data is received once and only once by each node,
b) data are routed on the shortest paths,
c) link loads are balanced.

Proof: The proof of the necessary condition is straightforward. Indeed when conditions a), b) and c) are all verified, the broadcast task is well balanced between all the nodes and then they all begin and end at the same time. Furthermore each node does just that is necessary. To prove the sufficient condition, suppose that one and only one of the conditions a), b), and c) is not verified.

- The condition a) is not verified. Necessarily, at some steps, at least one node receives data whose one piece is redundant. One of useful pieces of data which should be received instead of the redundant one has been necessarily sent by an other node. Thus there is a node which, at this step, has sent a larger piece of data which requires more time.
- The condition b) is not verified. Then more time is required to convey the data to their destination.
- The condition c) is not verified. Then overloaded links will take more time to transfer their data.

In any case, the data transfer requires more time; the resulting broadcast can not be optimal. At the best, it is ε-optimal.

We can deduce from this proposition that in an optimal all-to-all broadcast, the number of the data exchanges steps between adjacent nodes has to be equal to the diameter of the torus, at each step the amount of data to exchange has to be the same on each link and the data must be routed on the shortest paths.

## IV. MATHEMATICAL FOUNDATIONS

This section presents the mathematical properties used in the next sections to devise the conditions under which optimal all-to-all broadcast can be performed on a k-ary n-dimensional torus.

*A. Definitions*

In the remainder of the paper, we will use the following equivalent definition of a k-ary n-cube torus instead of the one given in the previous section.

*1) k-ary n-dimensional torus:* Equivalently to the previous definition, a k-ary n-dimensional torus can also be viewed as a network of $k^n$ nodes $x(x_1, x_2, \ldots, x_i, \ldots, x_n)$ such as $-\lfloor k/2 \rfloor \leq x_i \leq \lfloor k/2 \rfloor$) and two nodes are connected if and only if their addresses differ by 1 [modulo k] on one and only one digit.

*2) The reference node:* A reference node of a k-ary n-dimensional torus is a node from which the topological view of the k-ary n-cube is identical to the one from any other node.

By definition of the k-ary n-dimensional torus, it is obvious that any node can be the reference node. In the sequel, this will be the node which has all its coordinates equal to 0. It will be distinguished, as illustrated in Fig. 2, by a small black square.

*3) Boundary node:* A boundary node of a k-ary n-dimensional torus is a node which, observed from the reference node, possesses a wraparound link.

*4) Quadrant:* A quadrant of a k-ary n-dimensional torus is the set of the nodes whose the coordinates, in any dimension, are both positive or both negative.

For instance, in the k-ary n-dimensional of Fig. 2 there are four quadrants, one of which, the North-West quadrant, is constituted of the nodes $x(x_1, x_2)$ such as $x_1 \leq 0$ and $x_2 \geq 0$.

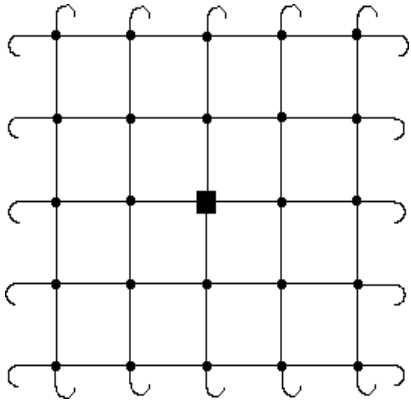

Fig. 2. 5-ary 2-dimensional torus

*B. Properties*

Let:

$C(m,p)$ be the number of combinations of p elements among m,

$\exp(q,m)$ be the exponent of the prime factor q in the decomposition of m,

$M_{(p,v)}(n, k, t)$ be the cardinal of the set of the nodes of a k-ary n-dimensional torus located at distance t from the reference node and having exactly p of their coordinates equal to v.

**Proposition 2:** Let $\mathbf{I}_{n,p}$ be the set of the subsets of the n first non null natural integer of cardinality p.

$$M_{(p,v)}(n,k,t)=G_{(p,v)}(n,k)*C(n,p)*N_{(p,v)}(n,k,t)$$

where:

$$G_{(p,v)}(n,k)= \begin{cases} 2^{n-p} & \text{if } v=0 \\ 2^n & \text{otherwise} \end{cases} \text{ if k is odd.}$$

$$G_{(p,v)}(n,k)= \begin{cases} 1 & \text{if } v=0 \\ 2^p & \text{otherwise} \end{cases} \text{ if k is even.}$$

$$N_{(p,v)}(n,k,t)= \begin{cases} |\{x:x_i \geq 0 \text{ and } \sum_{i \notin I} x_i = t-pv\}| & \text{if k is odd} \\ \sum_{0 \leq h \leq n-p} 2^{n-p-h}*|\{x \geq 0: \sum_{i \notin I}|x_i|=t-pv-h\lfloor k/2 \rfloor\}| & \text{otherwise.} \end{cases}$$

Proof: By definition,

$M_{(p,v)}(n,k,t)=|\bigcup_{I \in \mathbf{I}_{n,p}}\{x:|x_i| \leq \lfloor k/2 \rfloor, \sum_{1 \leq i \leq n}|x_i|=t$ and
$|x_i|=v$ if $i \in I\}|$

$=\sum_{I \in \mathbf{I}_{n,p}}|\{x:\sum_{i \notin I}|x_i|=t-pv\}|.$

Let $Q$ be the set of quadrants of the k-ary n-dimensional torus according to the reference node.

$M_{(p,v)}(n,k,t)=\sum_{I \in \mathbf{I}_{n,p}}\sum_{Q \in \mathcal{Q}}|\{x \in Q:\sum_{i \notin I}|x_i|=t-pv\}|$

$=|\mathbf{I}_{n,p}|*\sum_{Q \in \mathcal{Q}}|\{x \in Q:\sum_{i \notin I}|x_i|=t-pv\}|$

$=C(n,p)*\sum_{Q \in \mathcal{Q}}|\{x \in Q:\sum_{i \notin I}|x_i|=t-pv\}|.$

Two situations may arise according to the parity of k.

*Case 1:* k is odd. As illustrated in Fig. 2, all the quadrants are structurally identical. Then it is sufficient to reason on one of the quadrants. Let's consider, for instance, the quadrant:

$$Q=\{x:x_i=v \text{ if } i \in I \text{ and } x_i \geq 0 \text{ otherwise}\}.$$

$\sum_{Q \in \mathcal{Q}}|\{x \in Q:\sum_{i \notin I}|x_i|=t-pv\}|=|\mathcal{Q}|*|\{x \in Q:\sum_{i \notin I}|x_i|=t-pv\}|$

and then:

$$M_{(p,v)}(n,k,t)=C(n,p)*G_{(p,v)}(n,k)*N_{(p,v)}(n,k,t)$$

with:

$$G_{(p,v)}(n, k) = \begin{cases} 2^{n-p} & \text{if } v = 0 \\ 2^n & \text{otherwise,} \end{cases}$$

$N_{(p,v)}(n,k,t) = |\{x : x_i \geq 0 \text{ and } \sum_{i \notin I} x_i = t-pv\}|$.

*Case 2*: k is even. Again:

$$M_{(p,v)}(n,k,t) = C(n,p) * \sum_{Q \in \mathcal{Q}} |\{x \in Q : \sum_{i \notin I} |x_i| = t-pv\}|.$$

However, the quadrants are no more structurally identical. Indeed, they have not the same number of boundary nodes as illustrated in Fig. 3 for the 4-ary 2-dimensional torus.

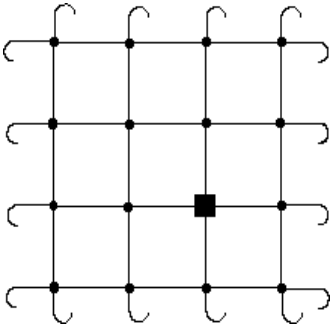

Fig. 3. 4-ary 2-dimensional torus

Then, according to set of the nodes x having h coordinates $x_i$, for $i \notin I$, equal to $\lfloor k/2 \rfloor$:

$$M_{(p,v)}(n,k,t) = C(n,p) * \sum_{0 \leq h \leq n} |\{x : \sum_{i \notin I} |x_i| = t-pv-h\lfloor k/2 \rfloor\}|.$$

Let's embed the k-ary n-dimensional torus in a (k+1)-ary n-dimensional torus. As illustrated in Fig. 4, the quadrants become structurally identical but with each boundary node appearing redundantly $2^h$ times for $0 \leq h \leq n-p$.

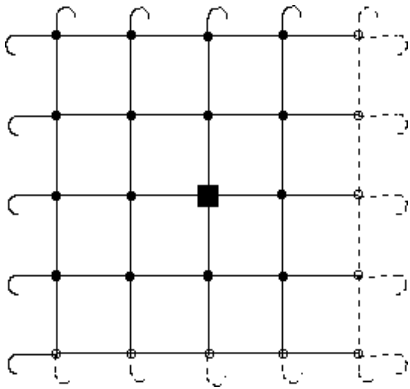

Fig. 4. 4-ary 2-dimensional torus embedded in 5-ary 2-dimension torus

Redundant boundary nodes are illustrated in Fig. 4 by white East (resp. South) boundary nodes which are duplicates of the West (resp. North) black boundary nodes while their induced redundant links are illustrated by the dotted links.

Then, from the odd-arity case:

$$M_{(p,v)}(n,k,t) = C(n,p) * \sum_{0 \leq h \leq n-p} G'_{(p,v)}(n,k)/2^h * |\{x : x_i \geq 0 \text{ and } \sum_{i \notin I} |x_i| = t-pv-h\lfloor k/2 \rfloor\}|$$
$$= C(n,p) * G'_{(p,v)}(n,k)/2^{n-p} * N_{(p,v)}(n,k,t).$$

From where the expected results:

$G'_{(p,v)}(n,k) = G_{(p,v)}(n, 2k'+1)$
$N_{(p,v)}(n,k,t) = \sum_{0 \leq h \leq n-p} 2^{n-p-h} * |\{x \geq 0 : \sum_{i \notin I} |x_i| = t-pv-h\lfloor k/2 \rfloor\}|$.

Let $(s_r, r \geq 0)$ be the recursive sequence whose general term is defined as:

$$s_r = \begin{cases} 1 & \text{if } r = 0 \\ (s_{r-1})^{q-1} \sigma_{r-1} & \text{otherwise} \end{cases}$$

where $(s_{r-1})^{q-1}$ stands for (q-1) times the terms of $s_{r-1}$ and $\sigma_{r-1}$ stands for the sequence $s_{r-1}$ whose last term is incremented by 1. For instance for q = 3 and r = 3 we obtain:

$s_0 = $ **1**
$s_1 = 1\ 1\ $**2**
$s_2 = 1\ 1\ $**2**$\ 1\ 1\ $**2**$\ 1\ 1\ $**3**
$s_3 = 1\ 1\ 2\ 1\ 1\ 2\ 1\ 1\ $**3**$\ 1\ 1\ 2\ 1\ 1\ 2\ 1\ 1\ $**3**$\ 1\ 1\ 2\ 1\ 1\ 2\ 1\ 1\ $**4**

Let $s_r(m)$ be the m-th term of the sequence $s_r$ and $m_i$ the i-th digit of m in the base q.

**Lemma 1:** $\sum_{1 \leq i \leq m} s_r(i) = \sum_{0 \leq i \leq \lfloor \log m \rfloor} \lfloor m/q^i \rfloor$.

Proof: We know from the decomposition of an integer in a base q that $m = \sum_{0 \leq i \leq \lfloor \log m \rfloor} m_i q^i$. Consequently, in accordance with the recursive nature of the sequence $s_r$ and the fact that $m_i < q$ it follows that:

$$\sum_{1 \leq i \leq m} s_r(i) = \sum_{0 \leq i \leq \lfloor \log m \rfloor} m_i \sum_{1 \leq j \leq qi} s_i(j).$$

From the definition of $s_i$, we know that:

$\sum_{1 \leq j \leq qi} s_i(j) = (q-1) * \sum_{1 \leq j \leq q(i-1)} s_{i-1}(j) + \sum_{1 \leq j \leq q(i-1)} \sigma_{i-1}(j)$
$= (q-1) * \sum_{1 \leq j \leq q(i-1)} s_{i-1}(j) + \sum_{1 \leq j \leq q(i-1)} s_{i-1}(j) + 1$
$= q * \sum_{1 \leq j \leq q(i-1)} s_{i-1}(j) + 1$

By iteration, using similar reasoning, we obtain:

$$\sum_{1 \leq j \leq qi} s_i(j) = q^2 * \sum_{1 \leq j \leq q(i-2)} s_{i-2}(j) + q + 1$$
$$= \sum_{0 \leq j \leq i} q^j.$$

Therefore:

$$\sum_{1 \leq i \leq m} s_r(i) = \sum_{0 \leq i \leq \lfloor \log m \rfloor} m_i \sum_{0 \leq j \leq i} q^j.$$

Let's consider the following triangular matrix organisation of the terms of the second member.

| | | | | | |
|---|---|---|---|---|---|
| i = 0 | $m_0$ | | | | |
| i = 1 | $m_1$ | $m_1 q$ | | | |
| i = 2 | $m_2$ | $m_2 q$ | $m_2 q^2$ | | |
| … | | | | | |
| i = $\log_q m$ | $m_{\log m}$ | $m_{\log m} q$ | $m_{\log m} q^2$ | … | $m_{\log m} q^{\log m}$ |

Summing the terms on the same diagonal we obtain:

$$\sum_{i \leq j \leq \lfloor \log m \rfloor - i} m_j q^j = \lfloor \sum_{0 \leq j \leq \lfloor \log m \rfloor} m_j q^j / q^i \rfloor = \lfloor m/q^i \rfloor$$

from where:

$$\sum_{1 \leq i \leq m} s_r(i) = \sum_{0 \leq i \leq \lfloor \log m \rfloor} \lfloor m/q^i \rfloor.$$

**Lemma 2:** $\exp(q, p!) = \sum_{1 \leq j \leq 1 + \lfloor \log(p/q) \rfloor} \lfloor p/q^j \rfloor.$

Proof: By definition,

$$\exp(q, p!) = \sum_{1 \leq m \leq p} \exp(q, m).$$

As the only values of m for which $\exp(q,m)$ is non null are the multiples of q, the above relation becomes:

$$\exp(q, p!) = \sum_{1 \leq j \leq \lfloor p/q \rfloor} \exp(q, j*q)$$

where:

$$\exp(q, j*q) = \begin{cases} 1 & \text{if } j < q \\ 1 + \exp(q, j) & \text{if } j \text{ is a power of } q \\ \exp(q, j*q \bmod (q^{1+\lfloor \log q \rfloor})) & \text{otherwise.} \end{cases}$$

In others words the sequence $(\exp(q,j*q), 1 \leq j \leq \lfloor p/q \rfloor)$ is straightforwardly a subsequence of $s_r$ with $r \geq \lfloor \log_q \lfloor p/q \rfloor \rfloor$. Then, from Lemma 1 it comes that:

$$\exp(q, p!) = \sum_{1 \leq j \leq \lfloor p/q \rfloor} s_r(j)$$
$$= \sum_{0 \leq j \leq \log \lfloor p/q \rfloor} \lfloor \lfloor p/q \rfloor / q^j \rfloor$$
$$= \sum_{0 \leq j \leq \log \lfloor p/q \rfloor} \lfloor p/q^{j+1} \rfloor$$
$$= \sum_{1 \leq j \leq 1 + \log \lfloor p/q \rfloor} \lfloor p/q^j \rfloor.$$

**Lemma 3:** Let $C_p = \sum_{1 \leq j \leq 1 + \lfloor \log((n-p)/q) \rfloor} \lfloor (n-p)/q^j \rfloor.$

$$C_p = (1/(q-1))\left((n-p) - \left(\sum_{0 \leq j \leq \alpha} n_j - \sum_{0 \leq j \leq \beta} p_j\right)\right) - \text{Card}\{1 \leq j \leq \beta: \sum_{0 \leq i \leq j-1} n_i q^i < \sum_{0 \leq i \leq j-1} p_i q^i\}.$$

Proof: Let's consider that:

$$n = \sum_{0 \leq i \leq \alpha} n_i q^i \text{ with } \alpha = \lfloor \log_q n \rfloor$$
$$p = \sum_{0 \leq i \leq \beta} p_i q^i \text{ with } \beta = \lfloor \log_q p \rfloor.$$

Then, as $n > p$:

$$n - p = \sum_{0 \leq i \leq \alpha} (n_i - p_i) q^i \lfloor (n-p)/q^j \rfloor = \sum_{j \leq i \leq \alpha} (n_i - p_i) q^{i-j} + c_j$$

where $c_j = \begin{cases} -1 & \text{if } \sum_{j \leq i \leq \alpha} (n_i - p_i) q^i < 0 \\ 0 & \text{otherwise.} \end{cases}$

$$C_p = \sum_{1 \leq j \leq \alpha} \lfloor (n-p)/q^j \rfloor$$
$$= \sum_{1 \leq j \leq \alpha} \sum_{j \leq i \leq \alpha} (n_i - p_i) q^{i-j} + \sum_{1 \leq j \leq \alpha} c_j$$
$$= \sum_{1 \leq j \leq \alpha} \sum_{j \leq i \leq \alpha} (n_i - p_i) q^{i-j} - \text{Card}\{1 \leq j \leq \beta: \sum_{0 \leq i \leq j-1} n_i q^i < \sum_{0 \leq i \leq j-1} p_i q^i\}$$
$$= \sum_{1 \leq j \leq \alpha} \sum_{j \leq i \leq \alpha} n_i q^i - \sum_{1 \leq j \leq \beta} \sum_{j \leq i \leq \beta} p_i q^i - \text{Card}\{1 \leq j \leq \beta: \sum_{0 \leq i \leq j-1} n_i q^i < \sum_{0 \leq i \leq j-1} p_i q^i\}.$$

As in the proof of Lemma 1, the terms of $\sum_{1 \leq j \leq \alpha} \sum_{j \leq i \leq \alpha} n_i q^i$ can be organized in the following triangular matrix:

| | | | | | |
|---|---|---|---|---|---|
| j = α | $n_\alpha$ | | | | |
| j = α-1 | $n_{\alpha-1}$ | $n_\alpha q$ | | | |
| j = α-2 | $n_{\alpha-2}$ | $n_{\alpha-1} q$ | $n_\alpha q^2$ | | |
| … | | | | | |
| j = 1 | $n_1$ | $n_2 q$ | $n_3 q^2$ | … | $n_\alpha q^{\alpha-1}$ |

Summing the terms on the same diagonal we obtain that:

$$\sum_{1 \leq j \leq \alpha} \sum_{j \leq i \leq \alpha} n_i q^i = \sum_{1 \leq j \leq \alpha} n_j \sum_{0 \leq i \leq j-1} q^i$$
$$= (1/(q-1)) \sum_{1 \leq j \leq \alpha} n_j (q^j - 1)$$
$$= (1/(q-1))\left(\sum_{1 \leq j \leq \alpha} n_j q^j - \sum_{1 \leq j \leq \alpha} n_j\right)$$
$$= (1/(q-1))\left(n - \sum_{0 \leq j \leq \alpha} n_j\right).$$

Similarly:

$$\sum_{1 \leq j \leq \beta} \sum_{j \leq i \leq \beta} p_i q^i = (1/(q-1))\left(p - \sum_{0 \leq j \leq \beta} p_j\right)$$

from where, we obtain the expected result.

## V. CHARACTERIZATION OF THE OPTIMALITY

The aim of this section is to characterize k-ary n-dimensional tori capable of optimal all-to-all broadcasts such as data are routed on shortest paths, received only once by each node, and link loads are balanced.

According to these constraints, given the incremental construction of the torus, at step t of the broadcast each data has to move in a descending construction order of the torus. This move can be realized according to several equivalent strategies. Such a strategy can be as simple as to move towards one of the nearest axis or the farthest axis of the torus with a smaller dimension. So let's choose the move-towards-one-of-the-nearest-axis strategy. This strategy partitions the nodes located at a given distance from the reference node into the classes of nodes at distance 0,1,… from one or more nearest axis. The nodes belonging to a same class can also be partitioned into classes of nodes having, in this order, exactly n,n-1,…,1 identical coordinates.

As each piece of data located at a distance t from the reference node and having exactly p of its coordinates equal to v can be routed to the reference node only on the p axis from which it is at distance v, a sufficient and necessary condition for an optimal all-to-all broadcast is that for any t, $M_{(p, v)}(n, k, t)$ must be divisible by the number of the incoming axis of any node which is equal to $2n$.

**Lemme 4:** $2n$ does not divide $N_{(p, v)}(n, k, t)$.

Proof: Let q be a prime factor of $2n$. From Proposition 2,

$$N_{(p, v)}(n,k,n(v+1)-p) = \begin{cases} 1 & \text{if k is odd} \\ 2^{n-p} & \text{otherwise.} \end{cases}$$

Then:
- if $q \neq 2$: $\exp(q, N_{(p, v)}(n,k,t)) < \exp(q, 2n)$,
- if $q = 2$: for k odd, $N_{(p, 0)}(n,k,n-p) = 1$ and again: $\exp(q, N_{(p, v)}(n,k,t)) < \exp(q, 2n)$.

**Lemma 5:** A necessary and sufficient condition for $2n$ to divide $G_{(p, v)}(n,k)C(n,p)$ is that:

$$\text{Card}\{j \leq \beta : \sum_{0 \leq i \leq j-1} n_i q^i < \sum_{0 \leq i \leq j-1} p_i q^i\} \geq \text{Card}\{j \leq \alpha : \sum_{0 \leq i \leq j-1} n_i q^i = 0\} - \exp(q, Q_{(p, v)}(n,k)) + \exp(q, 2)$$

where $\alpha = \lfloor \log_q n \rfloor$ and $\beta = \lfloor \log_q p \rfloor$, for any prime factor q of n and p.

Proof: Let's recall that $2n$ divides $G_{(p,v)}(n,k)C(n,p)$ means that for any prime factor q of $2n$ and $G_{(p, v)}(n,k)C(n,p)$ we have:

$$\exp(q,2) + \exp(q,n) \leq \exp(q, G_{(p, v)}(n,k)) + \exp(q, C(n,p))$$

According to the expression of $G_{(p,v)}(n,k)$ from Proposition 2, two situations may arise.

*Case 1*: $q \neq 2$. In this case we have $\exp(q,2) = 0$ and $\exp(q, G_{(p, v)}(n,k)) = 0$. Then the above inequality becomes:

$$\exp(q,n) \leq \exp(q, C(n,p))$$

Let's recall that:

$$C(n,p)/n = (n-1)(n-2)\ldots(n-p+1)/p!$$

and that, from Lemma 2:

$$\exp(q, p!) = \sum_{1 \leq j \leq 1 + \lfloor \log(p/q) \rfloor} \lfloor p/q^j \rfloor.$$

From the same lemma we also know that:

$$\exp(q,(n-1)(n-2)\ldots(n-p+1)) = \sum_{1 \leq j \leq 1 + \lfloor \log((n-1)/q) \rfloor} \lfloor (n-1)/q^j \rfloor - \sum_{1 \leq j \leq 1 + \lfloor \log((n-p)/q) \rfloor} \lfloor (n-p)/q^j \rfloor.$$

Indeed,

$$(n-1)(n-2)\ldots(n-p+1) = (n-1)!/(n-p)!.$$

We know from Lemma 3 that:

$$C_1 = (1/(q-1))((n-1) - (\sum_{0 \leq j \leq \alpha} n_j - 1)) - \text{Card}\{1 \leq j \leq \beta : \sum_{0 \leq i \leq j-1} n_i q^i < 1\}$$
$$= (1/(q-1))((n-1) - (\sum_{0 \leq j \leq \alpha} n_j - 1)) - \text{Card}\{1 \leq j \leq \beta : \sum_{0 \leq i \leq j-1} n_i q^i = 0\}.$$

Similarly,

$$\sum_{1 \leq j \leq 1 + \lfloor \log p/q \rfloor} \lfloor p/q^j \rfloor = (1/(q-1))(p - \sum_{0 \leq j \leq \beta} p_j).$$

By substituting to each term for its value in the above divisibility condition, we obtain the expected relation with the two last terms of the second member being equal to 0.

*Case 2*: $q = 2$. In this case, $\exp(q, 2) = 1$ and $\exp(q, G_{(p, v)}(n,k)) = 0$ or $n-p$. Then the divisibility condition becomes:

$$1 + \exp(q,n) \leq \exp(q, G_{(p, v)}(n,k)) + \exp(q, C(n,p)).$$

By a reasoning similar to the one used for the case where $q \neq 2$, we obtain the desired relation.

At this point of our characterization, we have to specify the values of n which satisfy the condition of Lemma 5. Again this depends on the values of q.

*Case 1*: $q \neq 2$. The question is to know if there is n such as:

$$\text{Card}\{j \leq \beta : \sum_{0 \leq i \leq j-1} n_i q^i < \sum_{0 \leq i \leq j-1} p_i q^i\} \geq \text{Card}\{j \leq \alpha : \sum_{0 \leq i \leq j-1} n_i q^i = 0\}$$

for any $p < n$.

The answer is definitively no. It suffices indeed to take p=q. Therefore the only values of n which are candidates are those which do not admit no other prime factor else q=2, that is, the values of n which are powers of 2.

*Case 2*: q=2. The question is to know if there is n, equal to a power of 2 and strictly greater than any p, such as:

$$\text{Card}\{j \leq \beta: \sum_{0 \leq i \leq j-1} n_i q^i < \sum_{0 \leq i \leq j-1} p_i q^i \} \geq \text{Card}\{j \leq \alpha: \sum_{0 \leq i \leq j-1} n_i q^i = 0\} - \exp(q, G_{(p,v)}(n,k)) + 1.$$

Two situations may arise according to the values of k.

*Case 2.1*: k is odd. As $G_{(p,v)}(n, k)$ may take different values we just have to verify the relation for the maximum of all values of the second member of the inequality which is attained for the minimum value of $\exp(q, G_{(p,v)}(n, k))$ that is $2^{n-p}$. Then we have:

$$\text{Card}\{j \leq \beta: \sum_{0 \leq i \leq j-1} n_i 2^i < \sum_{0 \leq i \leq j-1} p_i 2^i \} \geq \text{Card}\{j \leq \alpha: \sum_{0 \leq i \leq j-1} n_i 2^i = 0\} - n + p + 1$$

where $n=2^r$. Again let's consider the maximum of the second member of the inequality for all values of $p=2^r-2$. We obtain:

$$\text{Card}\{j \leq \alpha: \sum_{0 \leq i \leq j-1} n_i 2^i = 0\} - n + p + 1 = r - 1$$
$$\text{Card}\{j \leq \beta: \sum_{0 \leq i \leq j-1} n_i 2^i < \sum_{0 \leq i \leq j-1} p_i 2^i \} = r - 1$$

The inequality is then true.

*Case 2.2*: k is even. By the same reasoning as the case where k is odd, we have to verify the following inequality:

$$\text{Card}\{j \leq \beta: \sum_{0 \leq i \leq j-1} n_i 2^i < \sum_{0 \leq i \leq j-1} p_i 2^i \} \geq \text{Card}\{j \leq \alpha: \sum_{0 \leq i \leq j-1} n_i 2^i = 0\} + 1.$$

Again, from the case where q≠2, this relation can not be true. Indeed, it suffices to take p=2.

We can summarize this discussion by the following characterization which confirms the results obtained in [20] for k-ary 2-dimensional tori.

**Theorem:** A necessary and sufficient condition for an all-to-all broadcast to be optimal in a k-ary n-dimensional torus is that n is a power of 2 and k is odd.

## VI. CONCLUSION

This paper devised the conditions for optimal all-to-all broadcast in k-ary n-dimensional tori. Such a broadcast has to satisfy routing on the shortest paths while balancing the link-loads and minimizing the switching process at each node. The satisfaction of the balance of the link loads constraints imposes that the amount of data received on each link at each node has to be identical at each step of the broadcast; furthermore there must exist a partition of the data received at each node such as the cardinality of each element of the partition is divisible by the number of links.

The paper proves that such a partition can be built only for k-ary n-dimensional tori for which k is odd and n is a power of 2. In any other case, any all-to-all broadcast algorithm, at best, is ε-optimal but not optimal.

Then the objectives of the future works on this subject are double. On one side they will concern the study of the best all-to-all broadcast when k is even or n is not a power of 2. On the other side they will concern the study of the best suited switching processes in order to obtain efficient all-to-all broadcast, whatever are the arity and the dimension of the tori.

AUTHORS PROFILE

**Jean-Pierre Jung** received the PhD in computer science in 1983 at the University of Metz. From 1983 to 1994 he was Assistant Professor at the University of Metz. Since 1994 he is Professor at the Dpt. of Computer Science of the University of Metz where he teaches script and system programming. The research area of Prof. Jung concerns CAD and parallel and distributed systems.

**Ibrahima Sakho** received the PhD in applied mathematics in 1987 at the Institut National Polytechnique de Grenoble. From 1987 to 1992 he was at the Institute des Mathématiques Appliquées de Grenoble where he worked in the european supercomputer project Supernode, then from 1992 to 1997 at the École des Mines de St-Etienne where he was the head of the Parallel and Distributed Systems team. Since 1997 he is Professor at the University of Metz where he teaches computer architecture, parallel and distributed systems and making decision under uncertainty. The research of Prof. Sakho addresses the design of parallel and distributed algorithms and the algorithmic of the control of these systems.